\documentclass[twocolumn]{aastex631}

\begin{document}

\title{On the Excitation of Sloshing Oscillations in Hot Coronal Loops}

\author[0000-0001-9035-3245]{Arpit Kumar Shrivastav}
\affiliation{Southwest Research Institute, 1301 Walnut St, Suite 400, Boulder, CO 80302, USA}
\author[0000-0002-0735-4501]{S. Krishna Prasad}
\affiliation{Aryabhatta Research Institute of Observational Sciences (ARIES), Manora Peak, Nainital-263001, Uttarakhand, India}

\begin{abstract}

Observations of hot coronal loops have often shown the presence of compressive plasma oscillations. These oscillations are usually preceded by a flare-associated localised brightening in the vicinity of one of the footpoints of the coronal loop. The flare heating produces an impulsive injection of plasma into the coronal loop, along with a plasma perturbation that bounces back and forth between the two footpoints, termed as sloshing oscillations. While the association of these oscillations with flares is apparent, the pertinent connection between their excitation and the flare characteristics remains poorly understood. In this study, using data from the Atmospheric Imaging Assembly (AIA) onboard the Solar Dynamics Observatory (SDO), we present a unique set of observations, where the response of a single coronal loop due to repeated flares in the same active region is evident. Our analysis reveals that in three out of five flare events, sloshing oscillations are triggered, while in the other two events, only plasma flows are observed. By examining the duration of plasma injection associated with each flare, we find that sloshing oscillations occur when the injection duration is shorter than the travel time of the plasma perturbation across the loop. In contrast, no oscillations are observed when the injection duration is comparable to the travel time. These results reveal the dependence of sloshing oscillation excitation on the temporal characteristics of flares.

\end{abstract}


\section{Introduction} \label{sec:intro}

Spectroscopic observations from the Solar Ultraviolet Measurement of Emitted Radiation (SUMER) spectrometer \citep{1992Wilhelm_SUMER}, onboard the Solar and Heliospheric Observatory \citep{1995Domingo}, first revealed the presence of longitudinal intensity oscillations in coronal loops \citep{2002Wang}. These oscillations were prominently seen in hot coronal emission lines, with formation temperatures exceeding 6 MK, and have been clearly linked to flare related activity localized near one of the loop footpoints \citep{2003wang_b, 2005wang}. Due to characteristic physical properties such as propagation speeds matching local sound speeds and a quarter-period phase difference between intensity and Doppler velocity signals, these oscillations have been interpreted as standing slow magnetoacoustic waves \citep{2003wang_b}.

Later, the high-resolution imaging observations from the Atmospheric Imaging Assembly \citep[AIA;][]{2012Lemen_AIA} onboard the Solar Dynamics Observatory \citep[SDO;][]{2012Pesnell_SDO} revealed that plasma perturbations, associated with flares, propagate back and forth between the footpoints of coronal loops \citep{2013Kumar, kumar_x-ray_2015}. These events were termed ``sloshing oscillations" in different studies \citep{2016Real, 2019Nakariakov}. The properties of these oscillations closely resemble those of standing slow magnetoacoustic waves previously identified using SUMER observations \citep{2002Wang}. However, nonlinear MHD simulations suggest that sloshing oscillations can evolve into standing slow mode waves when higher harmonic components are dissipated, leaving the fundamental mode dominant \citep{2018Wang}. Supporting this theoretical prediction, recent AIA observations have provided the first evidence of such a transformation \citep{2021Krishna}.

The onset of sloshing oscillations in hot coronal loops can be attributed to impulsive heating events due to flares which generate ejected material with a hot plasma front ahead of it \citep{2013Kumar,2018Wang}. As the numerical simulations suggest, the critical factor that determines the generation of sloshing oscillations appears to be the duration of the impulsive perturbation relative to the loop sound crossing time \citep{2016Real}. Specifically, shorter impulsive perturbation relative to this characteristic time scale are more likely to induce sloshing oscillations. Despite this dependency seen in simulations, direct observational validation of the same is still lacking.

In addition to sloshing oscillations and standing slow waves in coronal loops, propagating periodic intensity disturbances are frequently observed in a variety of coronal structures, including active region fan loops \citep{2002DeMoortel, 2006Marsh, 2015Prasad, 2024Meadowcroft, 2025Tripathy} and polar plumes \citep{1997Ofman, 1998DeForest, 2012KrishnaPrasad, 2025Baweja}. These disturbances are often interpreted as propagating slow magnetoacoustic waves. However, this interpretation is not unique. Spectroscopic observations have revealed periodic blueward asymmetries in coronal emission line profiles, suggesting the presence of quasi-periodic high-speed upflows instead \citep{2010DePontieu, 2011Tian, 2012Tian}. Moreover, several studies have shown that waves and flows can be concurrent near the footpoints of coronal structures, with wave signatures often detectable over longer distances \citep{2015Samanta, 2016Bryans}. The close association between continuous plasma flows and propagating disturbances suggests that both phenomena may be driven by the same initiation mechanism, such as small-scale impulsive energy release \citep{2011Nishizuka,2012Wang}. Consistent with this interpretation, 3D MHD simulations indicate that slow magnetoacoustic waves and plasma flows at the base of active regions can be excited simultaneously by a common impulsive driver \citep{2012Ofman}. Despite these insights, the  occurrence of waves and flows in coronal loops and their connection to the underlying initiation mechanisms remain an open question in solar physics \citep{2021BanerjeeSSRv}.

\begin{figure*}
    \centering
    \includegraphics[width=\linewidth,clip,trim=0.1cm 0.7cm 0.5cm 0cm]{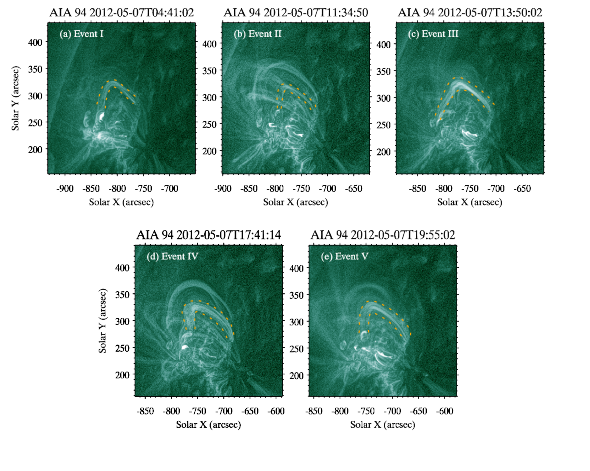}
    \caption{Context images of the observations used. The panels highlight the region of interest in the AIA 94 channel where flare related brightenings occurred near the footpoint of the same loop at different times. The orange dashed lines locate the position of the artificial slit representing the boundaries of the target loop structure at different times, as listed in Table \ref{tab:Table1}. An \href{https://drive.google.com/file/d/1HjHsssVVraIFZKd4MfHAu9tLuxBvajuU/view?usp=sharing}{animation} associated with the figure shows the temporal evolution of the loop response due to flares, including the excitation and evolution of sloshing oscillations along the loop.} 
    \label{fig:fig1_context}
\end{figure*}

\begin{figure*}
    \centering
    \includegraphics[width=\linewidth,trim=0.8cm 0cm 0.4cm 0.1cm]{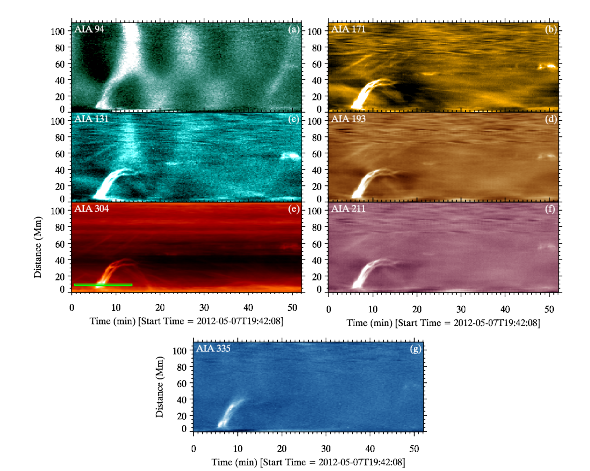}
    \caption{The time-distance maps derived from various EUV channels of AIA corresponding to the artificial slit marked in Figure \ref{fig:fig1_context} (e). The distance axis begins from the eastward footpoint of the loop. The horizontal green rectangular box, shown in panel (e) indicates the region chosen for further analysis and has a width of 5 pixels.}
    \label{fig:fig2_1900}
\end{figure*}

In this letter, we present new observational results that demonstrate a direct link between the initiation of sloshing oscillations in coronal loops and the timescale of flare-associated plasma injection, while also emphasizing the simultaneous generation of waves and flows driven by the same triggering events and their dependence on the characteristics of the energy release. The structure of the paper is as follows: Section \ref{sec:obs} outlines the observational data and the methodology employed to detect sloshing oscillations. Section \ref{analysis} details the analysis techniques used to characterize the oscillations and quantify the duration of plasma injections. Section \ref{sec:dis} offers a discussion of potential physical interpretations of our findings, followed by conclusions summarizing the main outcomes and implications of the study.

\section{Observations \& Data } \label{sec:obs}

The observations analyzed in this study focus on active region NOAA 11476, spanning the time period from 04:16 UT to 20:34 UT on May 07, 2012. The active region exhibited several small-scale flare events on the same day. One such event involved a C-class flare triggering sloshing oscillations, the characteristics of which were previously analyzed by \cite{2013Kumar} and \cite{2019Nakariakov}. Later, \cite{2021Krishna} performed the analysis by extending the same observation and demonstrated the transformation of sloshing oscillations into a standing slow-mode wave. 

We utilized datasets from SDO/AIA to study these small-scale flares and the associated plasma evolution in a coronal loop. AIA observes the Sun in seven extreme ultraviolet (EUV) channels, each sensitive to plasma at specific temperatures ranging from the chromosphere to the corona. The representative wavelengths of these channels are: 304 \AA\ (He II; $\sim$0.05 MK), 131 \AA\ (Fe VIII/Fe XXI; $\sim$0.4 MK/$\sim$10.0 MK), 171 \AA\ (Fe IX; $\sim$0.6 MK), 193 \AA\ (Fe XII/ Fe XXIV; $\sim$1.5 MK/20 MK), 211 \AA\ (Fe XIV; $\sim$2 MK), 335 \AA\ (Fe XVI; $\sim$2.5 MK), and 94 \AA\ (Fe XVIII; $\sim$7 MK) with the corresponding primary ions and their characteristic emission temperature listed in brackets \citep{2012Lemen_AIA}. However, the presence of multiple spectral lines in some passbands (e.g., 94 \AA, 131 \AA, 193 \AA) gives rise to distinct peaks in the temperature response function, and hence sensitivity to plasma at different temperatures within the same passband.  All the data were corrected for instrumental effects such as roll angle, plate scale, and spatial alignment between different wavelength channels using publicly available routines\footnote{https://robrutten.nl/rridl/sdolib/dircontent.html}. These routines convert the level 1 AIA data to science-grade level 1.5, yielding a final plate scale of 0.6 arcsec per pixel and a temporal cadence of 12 s.

The relevant AIA data at different time intervals, capturing five different episodes of small-scale flare related brightenings near the footpoint of the same coronal loop, are analyzed. A sample image of the target loop structure observed during each of these intervals is highlighted with orange dashed lines in Figure ~\ref{fig:fig1_context} as seen in the AIA 94 passband. Table \ref{tab:Table1} provides the start time of the footpoint brightening for different events. 

\begin{figure*}
    \centering
    \includegraphics[width=\linewidth,trim=8cm 0cm 6cm 1.5cm]{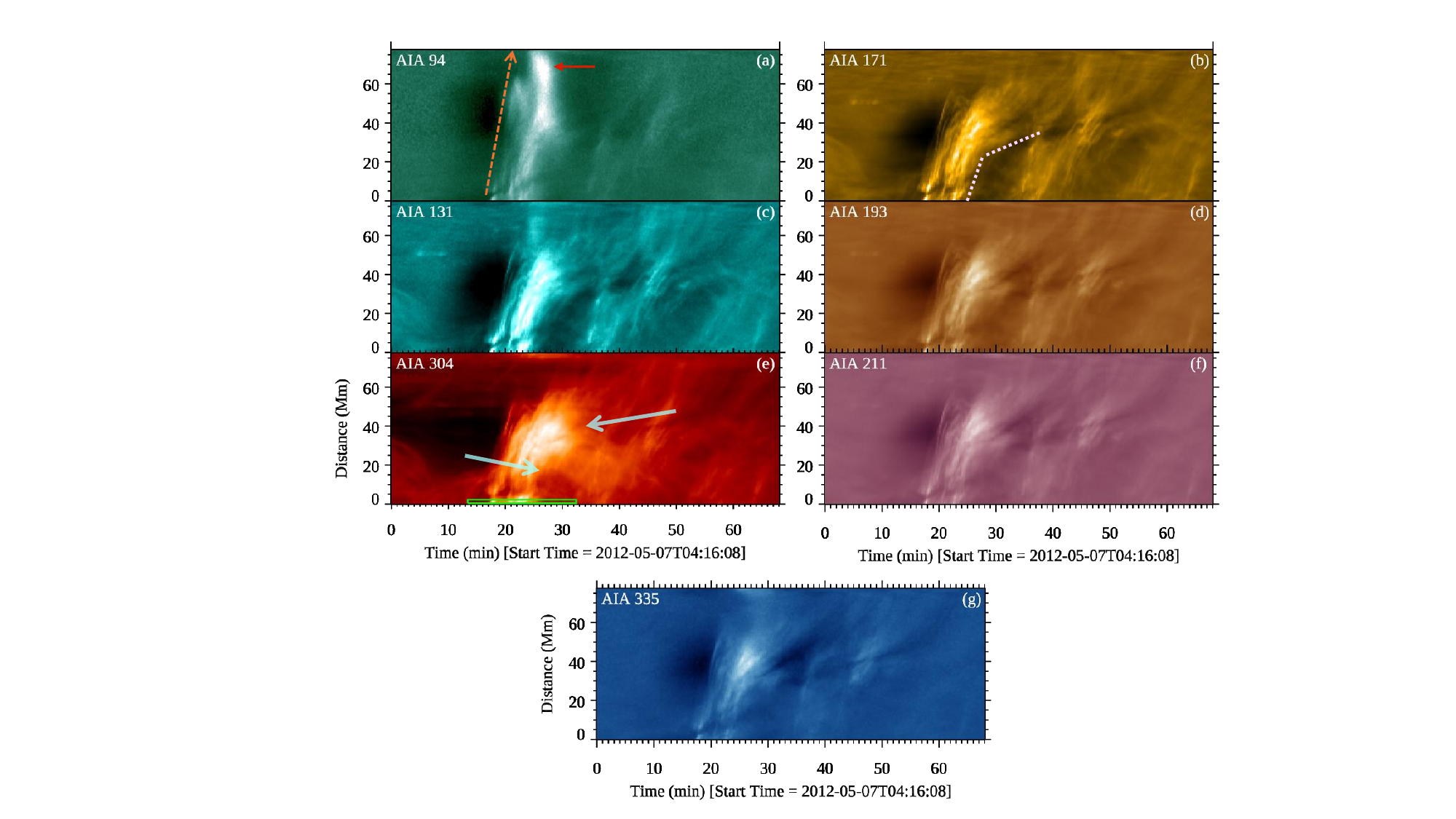}
    \caption{Similar to Figure \ref{fig:fig2_1900}, but corresponding to the artificial slit depicted in Figure \ref{fig:fig1_context}(a). The arrows and dashed lines shown in the different panels are described at the relevant locations in the text.}
    \label{fig:fig3_0400}
\end{figure*}

\section{Analysis \& Results} \label{analysis}

\subsection{Time–Distance Maps} \label{sec3.1}

Figure \ref{fig:fig1_context} shows snapshots of the active region NOAA 11476 as seen from the AIA 94 channel at five different times, highlighting a single loop that displays enhanced brightness following a flare-associated brightening near the eastward footpoint. Across different wavelength channels, the imaging sequence for each of these cases reveals injection of plasma into the loop in response to the flare. To analyze the spatio-temporal properties of these events (labeled with Roman numerals I to V in Table \ref{tab:Table1}), we first created a curved artificial slit by manually tracing the loop boundaries, shown by orange dashed lines in panels (a)–(e) of Figure \ref{fig:fig1_context}. We construct the time–distance maps by averaging the intensities transverse to the curvature of the slit from each image in the imaging sequence and by stacking the resultant intensity profiles together. To separate any patterns in the time–distance maps from the long-term background, we subtract a temporally smoothed image built using a running average of 24 minutes from the original time-distance maps. This background subtraction is done only to improve the visibility of short-scale fluctuations in the time–distance maps. The obtained time–distance maps for two specific cases are displayed in Figures \ref{fig:fig2_1900} and \ref{fig:fig3_0400}.

Figure \ref{fig:fig2_1900} (a)–(g) presents the time–distance maps in all seven EUV channels of AIA for event V (Table \ref{tab:Table1}), corresponding to the loop structure shown in Figure \ref{fig:fig1_context} (e). The repeated bright triangular pattern visible in the time–distance map of the AIA 94 channel reveals a compressive plasma perturbation moving along the loop from one foot point to the other and getting reflected multiple times before fading away. This motion is interpreted as sloshing oscillations, which in this case persisted for about three cycles. A weak but discernible signature of the same oscillation is also visible in the AIA 131 channel (Figure \ref{fig:fig2_1900} (c)), albeit for only about two cycles. In addition to the observed sloshing oscillation pattern, in all but the AIA 94 channel, we detect the presence of a single ``inverted U" pattern (see panels \ref{fig:fig2_1900}b–\ref{fig:fig2_1900}g) representing the rise and fall of plasma reaching only a fraction of the loop length, about 40 Mm distance from the eastward foot point. These plasma flows appear to initiate around the same time as the sloshing oscillations seen in AIA 94. The progressive decrease in brightness, especially during the falling phase, indicates that the plasma is rapidly cooling. The non-visibility of this plasma flow signature in the AIA 94 channel is perhaps because the oscillation signal is superimposed on that of the plasma flow, or because the flow temperature lies outside the sensitivity range of the AIA 94 passband. A visual comparison of the pattern visible in other channels favors the latter explanation. Note that the visibility of the flow pattern in the AIA 131 channel could also be because of its additional sensitivity to low temperature plasma. Taken together, the time-distance maps for event V reveal distinct signatures of both sloshing oscillations and plasma flows which originated around the same time, suggesting that both phenomena are excited by the same energy release event. Similar signatures highlighting the presence of both sloshing oscillations and plasma flows are found in the case of events III and IV (see Figures \ref{fig:appendix_fig2_1300} and \ref{fig:appendix_fig3_1700} in Appendix \ref{appendix:A}). Note that the sloshing oscillations detected in event IV correspond to the same event analyzed in earlier studies \citep{2013Kumar, 2021Krishna}.

Figure \ref{fig:fig3_0400}(a)–(g) presents the time–distance maps for event I, with the location of the artificial slit used for their construction marked in Figure \ref{fig:fig1_context}(a). None of these maps shows the repeated triangular pattern observed in the case of event V. However, in the AIA 94 channel, a bright ridge (or propagating front) extending from one foot point to the other is seen and shown by the dashed orange arrow in Figure \ref{fig:fig3_0400}(a). The enhanced brightening in the western section of the loop indicated by red arrow suggests that the plasma perturbation may have reflected from the western footpoint but failed to return to the eastern footpoint as seen in the time-distance map in Figure \ref{fig:fig3_0400}(a). This suggests the absence of fully developed sloshing oscillations spanning multiple cycles. All the other AIA channels display a plasma flow-like signature with apparent rising and falling motions. Notably, the flows here persist over a longer temporal duration than in event V, suggesting a more prolonged plasma injection in this case. Moreover, the distance from which the plasma starts to fall back is spread over a wider range as shown by arrows in Figure \ref{fig:fig3_0400}(e), while a portion of the plasma evidently crosses the apex and falls towards the other footpoint. Similar to event V, the falling part of the plasma is not as bright as the rising part, indicating rapid cooling. These flows, including the rising and falling part, are also simultaneously visible in the AIA 1600 \AA\ channel, suggesting the presence of a cooler plasma component. As shown by \cite{2019Simoes}, flare-excess emission in this passband is dominated by chromospheric and transition region lines with characteristic formation temperatures of approximately $10^{4.2}$--$10^{5.1}$ K. The 1600 \AA\ emission, together with the EUV observations in other AIA passbands, therefore suggests that these flows are multi-thermal. While comparing the time-distance map in AIA 94 with that from other channels, a notable deviation in the flow path may be seen, as indicated by the two dashed lines in Figure \ref{fig:fig3_0400}(b). These flows have an initial speed of 108$\pm$6 km  , which subsequently decreases to 39$\pm$2 km s$^{-1}$. 

Similarly, the time-distance maps corresponding to event II show no evidence of sloshing oscillations, nor even a single (or partial) reflection seen in event I (see Figure \ref{fig:appendix_fig1_1045}). Instead, we observe dominantly flow-like characteristics with rising and falling plasma visible across all channels except AIA 94. The pattern observed in AIA 94 is not very clear along the loop. Additionally, in this case, too, the plasma flows exhibit a longer temporal extent than in events where fully developed sloshing oscillations are observed. The time–distance maps for event II are provided in Figure \ref{fig:appendix_fig1_1045} in Appendix \ref{appendix:A}.

Overall, the time–distance maps for all five events indicate that fully developed sloshing oscillations, spanning multiple cycles, are triggered in only three cases. Although the plasma flows are observed in all cases, in the remaining two cases, the flow behavior is predominant. To further investigate and quantitatively assess the difference between the events, we estimate the associated duration of plasma injection into the loop, as discussed in the next section.    

\begin{figure*}
    \centering
    \includegraphics[width=\linewidth,trim=0cm 0cm 0cm 0cm]{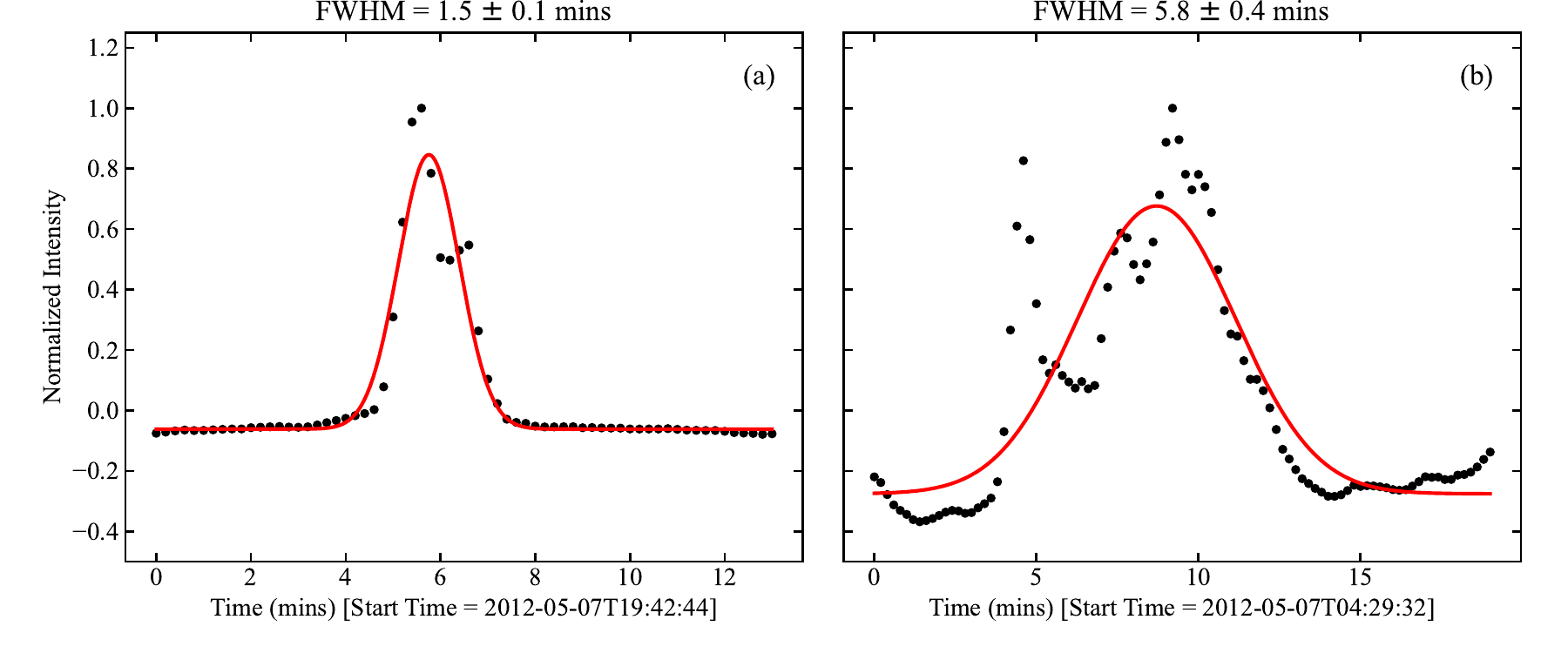}
    \caption{The scatter points in panels (a) and (b) represent the temporal variation of normalized intensity corresponding to the green boxes shown in Figure \ref{fig:fig2_1900}(e) and Figure \ref{fig:fig3_0400}(e). The red curve shows the fitted Gaussian function on these scatter points, delineating the overall envelope of enhancement.}
    \label{fig:fig4_plasma_injection_time}
\end{figure*}

\subsection{Duration of plasma injection}

\cite{2005wang} analyzed spectral line profiles associated with several slow mode oscillation events triggered by flares and found that, during the initial phase of the oscillation, the line profiles exhibited two distinct components: one corresponding to the background plasma and the other shifted component indicating high-speed flows (on the order of 100–300 km/s), likely representing the passage of impulsively injected hot plasma. These initial high-speed flows, accompanied by a Doppler velocity pulse, lasted for less than half a wave period, providing useful constraints on the heating timescale and serving as a proxy for the duration of impulsive energy release during the flare. However, accurately determining the duration of the second (flow) component is often challenging, as the spectral profiles may not always show a clear two-component structure suitable for double-Gaussian fitting.

In contrast, imaging observations offer a more direct means of constraining the duration of plasma flows. This fact assumes that the propagating plasma signatures trace genuine mass motion rather than a purely thermal conduction front. This assumption is reasonable considering the sound speeds could be of the order of 400-500 km s$^{-1}$ and the measured flow speeds are sub-sonic (Section \ref{sec3.1}). Since plasma flows are observed in all events, here, we determine the relevant time interval during which plasma is injected into the loop using the time–distance maps. For this, we place 5 pixel wide green boxes along the time axis at a fixed position in distance, near the footpoint, as shown in panel (e) of Figures \ref{fig:fig2_1900} and \ref{fig:fig3_0400}. The corresponding time-dependent normalized average intensities obtained from the green boxes are shown in Figure \ref{fig:fig4_plasma_injection_time}. Panels (a) and (b) display the results for events V and I, respectively. As may be recalled, event V exhibits sloshing oscillations that persist for three periods, whereas event I displays barely one single reflection (see Figures \ref{fig:fig2_1900} and \ref{fig:fig3_0400}). The intensity profiles in these two cases differ significantly. In event V, a single sharp peak can be seen (Figure \ref{fig:fig4_plasma_injection_time} (a)), while in event I, multiple peaks are observed (Figure \ref{fig:fig4_plasma_injection_time} (b)). These multiple peaks reflect the intermittent nature of the injected plasma, suggesting that the underlying driver itself may also be intermittent in nature. Nevertheless, considering the overall envelope of the enhancement as associated with a single flare-related driving episode (in agreement with the observed behavior in time-distance maps), we determine the duration of plasma injection into the loop from the full-width-at-half-maximum (FWHM) of a Gaussian fitted to the obtained intensity profiles. For event V, this value is found to be 1.5 minutes, whereas for event I, it is 5.8 minutes. These values suggest that the plasma injection time scale is relatively short in the case where sloshing oscillations are excited, compared to the case where no significant oscillations are observed. A similar trend is found in events III and IV, where sloshing oscillations are triggered, in contrast to event II, where only flow signatures with no apparent oscillations are detected (see Figure \ref{fig:appendix_fig1_1045}, \ref{fig:appendix_fig2_1300}, \ref{fig:appendix_fig3_1700} and in Appendix \ref{appendix:A}). The obtained intensity profiles and the corresponding fitted Gaussians for all the other events are shown in Figure \ref{fig:appendix_fig4_flare_duration} in the Appendix. The respective durations are listed in Table~\ref{tab:Table1}.

For comparison, we also estimate the travel time of the plasma perturbation from one end of the loop to the other. We do this by calculating the time difference between the peaks in intensity (determined from a Gaussian fit to the local intensity profile) near the foot points obtained from successive reflections in time-distance maps. Note that this estimate is not impacted by the plane-of-the-sky projection of the loop since there is no measurement of distance involved. The obtained travel times are 5.4, 4.7, and 6.2 minutes for events III, IV, and V, respectively. Since sloshing oscillations are not observed in events I and II, these values could not be estimated. However, because all the events occur within the same loop structure, assuming a similar travel time for these two cases implies that sloshing oscillations are likely to be excited when the plasma injection duration is shorter than the travel time of the plasma perturbation along the loop. In contrast, when the injection duration is comparable to the travel time between footpoints, no significant oscillations may be observed. 

\begin{figure*}
    \centering
    \includegraphics[width=\linewidth,trim=0cm 0.0cm 0cm 0.5cm]{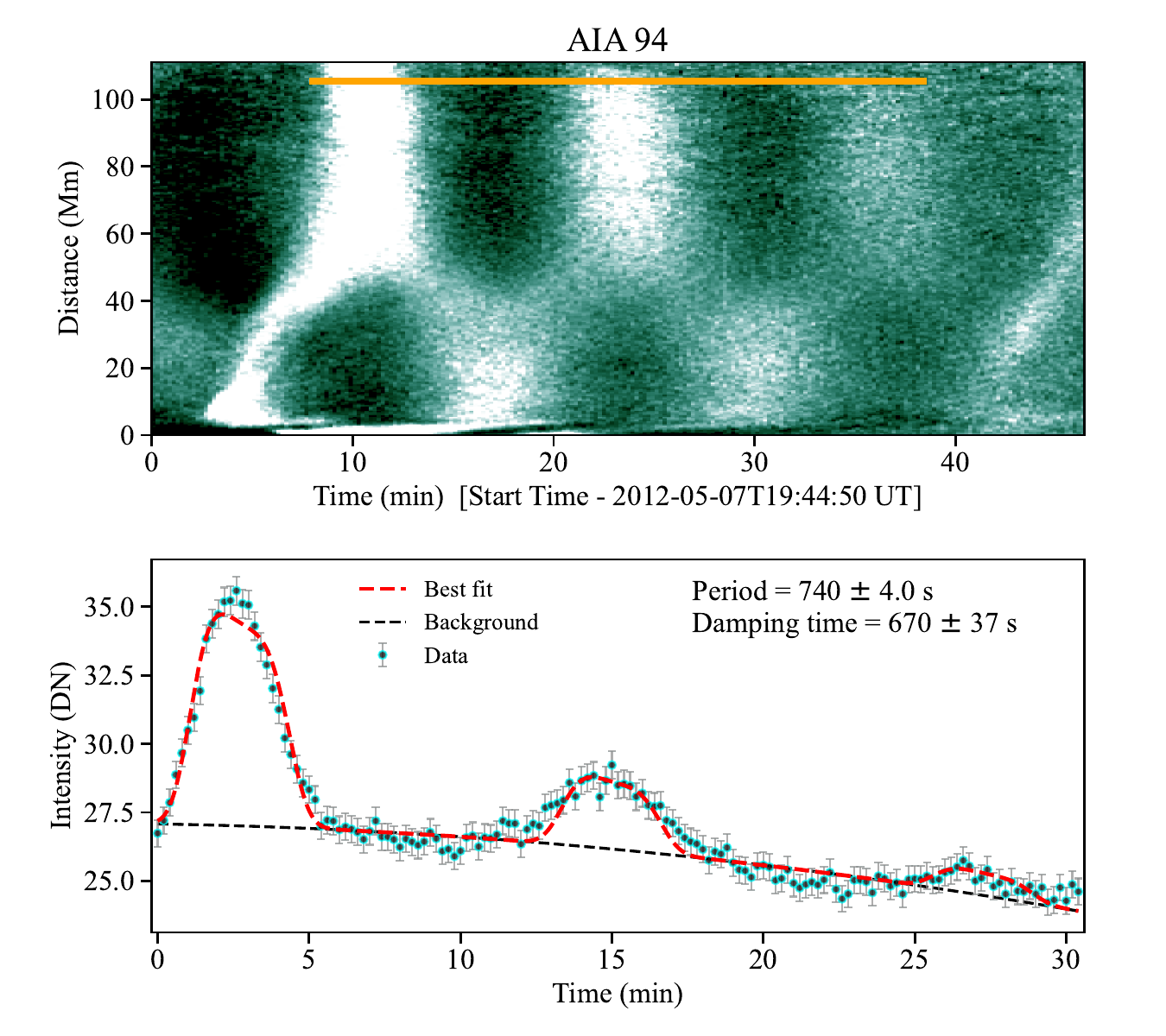}
    \caption{Estimate of oscillation parameters. The upper panel displays the time–distance map along the artificial slit marked in Figure \ref{fig:fig1_context}(e) for event V. The orange horizontal line indicates the spatial region over which the intensity time series, shown in the lower panel, is extracted. The data points represent intensity values obtained from the unprocessed time–distance map. The dashed red line denotes the fitted function following Equation \ref{eq:equation1}.}
    \label{fig:fig5_fitted_oscillations}
\end{figure*}

\begin{table*}
    \centering
    \small
    \renewcommand{\arraystretch}{1.25}
    \setlength{\tabcolsep}{3pt}

    \begin{tabular*}{\textwidth}
    {@{\extracolsep{\fill}}lccccccccc@{}}
    \hline
    \noalign{\vskip 4pt}
    Event &
    \shortstack{Start time\\(UT)} &
    \shortstack{Flare\\type} &
    \shortstack{Sloshing\\response} &
    \shortstack{$T_{\rm inj}$\\(min)} &
    \shortstack{$L$\\(Mm)} &
    \shortstack{$T_{\rm tr}$\\(min)} &
    \shortstack{$P$\\(s)} &
    \shortstack{$V_{\rm ph}$\\(km s$^{-1}$)} &
    \shortstack{$\tau$\\(s)} \\[4pt]
    \hline 
    \noalign{\vskip 4pt}

    I &
    04:33 &
    Mi/Br &
    \shortstack{Single\\reflection} &
    $5.8 \pm 0.4$ &
    $166.2 \pm 3.4$ &
    -- &
    -- &
    -- &
    -- \\[2pt]

    II &
    11:25 &
    Mi/Br &
    No &
    $6.1 \pm 0.6$ &
    $166.1 \pm 2.7$ &
    -- &
    -- &
    -- &
    -- \\[2pt]

    III &
    13:30 &
    Mi/Br &
    Yes &
    $2.6 \pm 0.1$ &
    $166.9 \pm 4.5$ &
    $5.4 \pm 0.1$ &
    $660 \pm 5$ &
    $505 \pm 26$ &
    $470 \pm 13$ \\[2pt]

    IV &
    17:29 &
    C7.4 &
    Yes &
    $2.2 \pm 0.2$ &
    $167.3 \pm 2.9$ &
    $4.7 \pm 0.1$ &
    $617 \pm 3$ &
    $542 \pm 10$ &
    $830 \pm 50$ \\[2pt]

    V &
    19:47 &
    Mi/Br &
    Yes &
    $1.5 \pm 0.1$ &
    $168.3 \pm 4.2$ &
    $6.2 \pm 0.1$ &
    $740 \pm 4$ &
    $454 \pm 17$ &
    $670 \pm 37$ \\[2pt]

    \hline
    \end{tabular*}

    \caption{Summary of the flare events and the derived plasma injection and oscillation parameters. Here, Mi/Br denotes a microflare or flare-associated brightening for the cases where flare class is difficult to identify. $T_{\rm inj}$ is the duration of the apparent plasma injection signature estimated from the FWHM of the fitted Gaussian profiles from x-t maps, $L$ is the deprojected loop length, $T_{\rm tr}$ is the travel time of the plasma perturbation between the two loop footpoints, $P$ is the oscillation period, $V_{\rm ph}=2L/P$ is the estimated phase speed, and $\tau$ is the exponential damping time. The symbol ``--'' indicates that the corresponding parameter could not be measured because fully developed sloshing oscillations were not detected.}
    \label{tab:Table1}
\end{table*}

\subsection{Properties of Oscillations}

Since sloshing oscillations are triggered in three separate instances within the same loop, we investigate how the oscillation properties vary across these events and how they relate to the underlying driving mechanisms. To analyze the oscillation characteristics, we extract intensity time series near the loop footpoints (orange line in Figure \ref{fig:fig5_fitted_oscillations}), which provides a more reliable representation of sloshing oscillation behavior \citep{2021Krishna}.  The light curve of the oscillation is then fitted using the functional form proposed by \cite{2021Krishna} and modified by \cite{2026Paliwal}, as given below:

\begin{equation}
I(t) = A_0 \, e^{-t/\tau} \, e^{-x_0^2} + B_0 t^{2} + C_0 t +D
\label{eq:equation1}
\end{equation}

\noindent where 
\[
x_0 = \frac{1}{2\sigma_n} \left( 1 - \cos\left( \frac{2\pi t}{P} + \phi \right) \right).
\]

\noindent Here, \( t \) denotes time, \( P \) is the oscillation period, \( \tau \) represents the damping time, and \( \sigma_n = \sigma / L \) corresponds to the normalized spatial width of the perturbation. The constants \( A_0 \) and \( \phi \) represent the amplitude and phase, respectively, while \( B_0 \), \( C_0 \) and D represents constants in the background trends. An example of the fitted oscillation profile is shown in Figure~\ref{fig:fig5_fitted_oscillations}, with the corresponding period and damping time listed in the plot. These parameters obtained for all three events where oscillations are observed, are summarized in Table~\ref{tab:Table1}. 

Events III and IV show a marginal difference in the oscillation period, whereas event V shows a noticeably longer period. The variation in periods may be attributed to differences in loop temperature. As demonstrated by \cite{2021Krishna}, during the transition from the sloshing phase to the standing phase, a decrease in plasma temperature can lead to a longer period due to the associated reduction in sound speed. The respective damping times show significant differences between all three events. Event III exhibits the fastest damping, with a damping time of 470$\pm$13 s, while Event IV shows the slowest, with a damping time of 830$\pm$50 s. Note that the variation in damping time is influenced by how effectively non-ideal effects, such as thermal conduction, compressive viscosity and wave induced thermal misbalance, operate within the loop during the oscillations \citep{2018Wang,2019Kolotkov,2021Krishna}. The substantial differences between the events may imply the role of flare in shaping the local physical conditions. There is no clear relation between the plasma injection duration and either the oscillation period or the damping time. However, the current sample size is too small to draw robust conclusions about their relationship, if such relation exist at all. Future statistical studies involving a larger number of sloshing oscillation events may be useful to determine this.

In addition to the oscillation parameters, we estimated the deprojected length of our target loop structure under the assumption of a circular loop geometry, following \citet{2002Aschwanden}. This was performed independently for each event, yielding similar values around 166 to 168 Mm. Considering the estimated loop length and oscillation period, we calculate the corresponding phase speed, $V_{ph}$, using \[V_{ph} = 2L/P.\] The estimated loop lengths and phase speeds are shown in Table~\ref{tab:Table1} for different cases. The phase speed values for events III, IV, and V are 505, 542, and 454 km s$^{-1}$, respectively. Previous studies using AIA have reported phase speeds of sloshing oscillations ranging from 397 to 853 km s$^{-1}$ \citep{2013Kumar, kumar_x-ray_2015, 2021Krishna, 2024Karakotov, 2026Paliwal} and the values obtained here fall within this range.

\section{Discussion \& Conclusions}\label{sec:dis}
In this study, we explore the response of a coronal loop within a flaring active region that is subjected to multiple flares in a single day. More specifically, we identify 5 flaring events that occurred over a span of $\approx$15 hours and investigate their role and efficacy in the excitation of sloshing oscillations in our target coronal loop. We utilise multi-channel imaging observations from SDO/AIA to achieve this objective. Based on time-distance analysis, we find that in three out of five cases, fully developed sloshing oscillations are produced. In one of the remaining two cases, the perturbation appears to travel to the opposite footpoint and undergoes a single reflection, but it quickly disappears before making it back to the starting point, resulting in an incomplete cycle. No signatures of oscillations are detected in the final case. 

Previous studies confirm that high velocity chromospheric evaporation flows are co-temporal with impulsive hard X-ray emission, indicative of energy deposition in the chromosphere due to impulsive energy release \citep{2009Milligan,2015Tian,2017Li}. Consequently, the duration of plasma injection inferred from EUV time–distance maps is expected to correlate with the underlying energy deposition timescale. However, the two timescales need not be exactly equal, since the injection is a hydrodynamic response to the energy deposition rather than a direct measure of it. It nonetheless provides a useful observational constraint on the relative timescale of the driver. Our analysis reveals that this plasma injection duration is an indicator of whether sloshing oscillations are excited; specifically, injections that are short relative to the loop’s sound crossing time are significantly more likely to produce coherent, long-lived oscillations. These observational results provide strong support for the numerical findings of \cite{2016Real}, who demonstrated through one-dimensional hydrodynamic simulations that sloshing oscillations are triggered only when the heat pulse duration is shorter than the loop’s sound crossing time.

By examining cases where sloshing oscillations are observed, it can be seen that flows and oscillations commence at the same time (see Figure \ref{fig:fig2_1900} for an example). They travel together for a short period before the flows recede and fall back while the oscillations continue. This implies both the bulk flows and the oscillatory perturbations originate from the same driving mechanism: the impulsive energy release as seen in our cases. Specifically, the impulsive heating initiates rapid chromospheric evolution and subsequent evaporation, injecting mass and momentum into the loop. However, if the presence of plasma flows and sloshing oscillations is dependent primarily on the energy deposition timescale or the duration of plasma injection, then their excitation should not depend strongly on the location of the heating. Similar signatures may also be observed when energy is deposited near the loop top and subsequently redistributed along the loop by efficient field aligned thermal conduction.

The flows only reach a fraction of the loop length in several cases presented here. This behavior can also be related to the low temperature of the flowing plasma, since the sound speed itself is lower at these temperatures. In addition, numerical simulations have shown that evaporating chromospheric plasma may decelerate while compressing the pre-existing coronal plasma and, as a result, may only partially fill the loop without reaching the apex \citep{2023Cozzo}. Also, there could be asymmetric effects in play as discussed in \cite{2023Cozzo}. 

However, in the case of event I (see Figure \ref{fig:fig3_0400}), the flows travel much larger distances, resulting in a portion of the plasma crossing the apex and falling towards the other footpoint. The wave signatures are very weak in this case, barely displaying one full cycle of oscillation. Combining the behavior of flows and waves seen in all events, one may conjecture that the energy deposited by the flare is partitioned between compressive waves and bulk mass transport. This energy partitioning appears to be governed by the rate (or duration) of energy deposition near the footpoint. As suggested by \cite{2016Real}, a short heat pulse can produce a pressure dip that drives sloshing oscillations, whereas prolonged heating allows the pressure to equilibrate along the loop, preventing the formation of such a dip. Based purely on the observed behavior, our scenario is also closer to a flow-pulse driver exciting compressive waves, similar to the MHD simulations reported by \cite{2013Wang}. These findings also align with MHD models that demonstrate the generation of waves and flows due to impulsive heating \citep{2012Ofman}, but highlight the need for detailed investigation to quantify how flare energy is channeled into these distinct dynamical components. More observational studies are also required in this direction.

\begin{acknowledgments}
\textit{Acknowledgments:} The authors thank the anonymous referee for constructive suggestions which improved the manuscript. AKS gratefully acknowledges SwRI for providing him necessary facilities to carry out this work. SKP is grateful to SERB/ANRF for a startup research grant (No. SRG/2023/002623) and ISRO for a RESPOND grant (No. ISRO/RES/2/453/25-26).
\end{acknowledgments}

\bibliography{bib_sloshing_trigger}{}

@ARTICLE{2013Kumar,
       author = {{Kumar}, Pankaj and {Innes}, D.~E. and {Inhester}, B.},
        title = "{Solar Dynamics Observatory/Atmospheric Imaging Assembly Observations of a Reflecting Longitudinal Wave in a Coronal Loop}",
      journal = {\apjl},
         year = 2013,
        month = dec,
       volume = {779},
       number = {1},
          eid = {L7},
        pages = {L7},
          doi = {10.1088/2041-8205/779/1/L7},
archivePrefix = {arXiv},
       eprint = {1409.3896},
 primaryClass = {astro-ph.SR},
       adsurl = {https://ui.adsabs.harvard.edu/abs/2013ApJ...779L...7K}
}

@ARTICLE{2002Wang,
       author = {{Wang}, Tongjiang and {Solanki}, S.~K. and {Curdt}, W. and {Innes}, D.~E. and {Dammasch}, I.~E.},
        title = "{Doppler Shift Oscillations of Hot Solar Coronal Plasma Seen by SUMER: A Signature of Loop Oscillations?}",
      journal = {\apjl},
         year = 2002,
        month = jul,
       volume = {574},
       number = {1},
        pages = {L101-L104},
          doi = {10.1086/342189},
       adsurl = {https://ui.adsabs.harvard.edu/abs/2002ApJ...574L.101W}
}

@ARTICLE{2018Wang,
       author = {{Wang}, Tongjiang and {Ofman}, Leon and {Sun}, Xudong and {Solanki}, Sami K. and {Davila}, Joseph M.},
        title = "{Effect of Transport Coefficients on Excitation of Flare-induced Standing Slow-mode Waves in Coronal Loops}",
      journal = {\apj},
         year = 2018,
        month = jun,
       volume = {860},
       number = {2},
          eid = {107},
        pages = {107},
          doi = {10.3847/1538-4357/aac38a},
archivePrefix = {arXiv},
       eprint = {1805.03282},
 primaryClass = {astro-ph.SR},
       adsurl = {https://ui.adsabs.harvard.edu/abs/2018ApJ...860..107W}
}

@ARTICLE{2021Krishna,
       author = {{Krishna Prasad}, S. and {Van Doorsselaere}, T.},
        title = "{Compressive Oscillations in Hot Coronal Loops: Are Sloshing Oscillations and Standing Slow Waves Independent?}",
      journal = {\apj},
         year = 2021,
        month = jun,
       volume = {914},
       number = {2},
          eid = {81},
        pages = {81},
          doi = {10.3847/1538-4357/abfb01},
archivePrefix = {arXiv},
       eprint = {2104.12038},
 primaryClass = {astro-ph.SR},
       adsurl = {https://ui.adsabs.harvard.edu/abs/2021ApJ...914...81K}
}

@ARTICLE{2003wang_b,
       author = {{Wang}, T.~J. and {Solanki}, S.~K. and {Curdt}, W. and {Innes}, D.~E. and {Dammasch}, I.~E. and {Kliem}, B.},
        title = "{Hot coronal loop oscillations observed with SUMER: Examples and statistics}",
      journal = {\aap},
         year = 2003,
        month = aug,
       volume = {406},
        pages = {1105-1121},
          doi = {10.1051/0004-6361:20030858},
       adsurl = {https://ui.adsabs.harvard.edu/abs/2003A&A...406.1105W}
}

@ARTICLE{2005wang,
       author = {{Wang}, T.~J. and {Solanki}, S.~K. and {Innes}, D.~E. and {Curdt}, W.},
        title = "{Initiation of hot coronal loop oscillations: Spectral features}",
      journal = {\aap},
         year = 2005,
        month = may,
       volume = {435},
       number = {2},
        pages = {753-764},
          doi = {10.1051/0004-6361:20052680},
       adsurl = {https://ui.adsabs.harvard.edu/abs/2005A&A...435..753W}
}

@ARTICLE{2016Real,
       author = {{Reale}, F.},
        title = "{Plasma Sloshing in Pulse-heated Solar and Stellar Coronal Loops}",
      journal = {\apjl},
         year = 2016,
        month = aug,
       volume = {826},
       number = {2},
          eid = {L20},
        pages = {L20},
          doi = {10.3847/2041-8205/826/2/L20},
archivePrefix = {arXiv},
       eprint = {1607.01329},
 primaryClass = {astro-ph.SR},
       adsurl = {https://ui.adsabs.harvard.edu/abs/2016ApJ...826L..20R}
}

@MISC{1992Wilhelm_SUMER,
       author = {{Wilhelm}, K. and {Axford}, W.~I. and {Curdt}, W. and {Gabriel}, A.~H. and {Grewing}, M. and {Huber}, M.~C.~E. and {Jordan}, S.~D. and {K{\"u}hne}, M. and {Lemaire}, P. and {Marsch}, E. and {Poland}, A.~I. and {Richter}, A.~K. and {Thomas}, R.~J. and {Timothy}, J.~G. and {Vial}, J.~C.},
        title = "{``SUMER'' - Solar Ultraviolet Measurements of Emitted Radiation.}",
 howpublished = {In ESA, Environment Observation and Climate Modelling Through International Space Projects. Space Sciences with Particular Emphasis on High-Energy Astrophysics p 225-226 (SEE N93-23878 08-88)},
         year = 1992,
        month = jul,
        pages = {225-226},
       adsurl = {https://ui.adsabs.harvard.edu/abs/1992eocm.rept..225W}
}

@ARTICLE{1995Domingo,
       author = {{Domingo}, V. and {Fleck}, B. and {Poland}, A.~I.},
        title = "{SOHO: The Solar and Heliospheric Observatory}",
      journal = {\ssr},
         year = 1995,
        month = apr,
       volume = {72},
       number = {1-2},
        pages = {81-84},
          doi = {10.1007/BF00768758},
       adsurl = {https://ui.adsabs.harvard.edu/abs/1995SSRv...72...81D}
}

@ARTICLE{2012Pesnell_SDO,
       author = {{Pesnell}, W. Dean and {Thompson}, B.~J. and {Chamberlin}, P.~C.},
        title = "{The Solar Dynamics Observatory (SDO)}",
      journal = {\solphys},
         year = 2012,
        month = jan,
       volume = {275},
       number = {1-2},
        pages = {3-15},
          doi = {10.1007/s11207-011-9841-3},
       adsurl = {https://ui.adsabs.harvard.edu/abs/2012SoPh..275....3P}
}

@ARTICLE{2012Lemen_AIA,
author="Lemen, James R.
and Title, Alan M.
and Akin, David J.
and Boerner, Paul F.
and Chou, Catherine
and Drake, Jerry F.
and Duncan, Dexter W.
and Edwards, Christopher G.
and Friedlaender, Frank M.
and Heyman, Gary F.
and Hurlburt, Neal E.
and Katz, Noah L.
and Kushner, Gary D.
and Levay, Michael
and Lindgren, Russell W.
and Mathur, Dnyanesh P.
and McFeaters, Edward L.
and Mitchell, Sarah
and Rehse, Roger A.
and Schrijver, Carolus J.
and Springer, Larry A.
and Stern, Robert A.
and Tarbell, Theodore D.
and Wuelser, Jean-Pierre
and Wolfson, C. Jacob
and Yanari, Carl
and Bookbinder, Jay A.
and Cheimets, Peter N.
and Caldwell, David
and Deluca, Edward E.
and Gates, Richard
and Golub, Leon
and Park, Sang
and Podgorski, William A.
and Bush, Rock I.
and Scherrer, Philip H.
and Gummin, Mark A.
and Smith, Peter
and Auker, Gary
and Jerram, Paul
and Pool, Peter
and Soufli, Regina
and Windt, David L.
and Beardsley, Sarah
and Clapp, Matthew
and Lang, James
and Waltham, Nicholas",
editor="Chamberlin, Phillip
and Pesnell, William Dean
and Thompson, Barbara",
year="2012",
journal = {\solphys},
volume = {275},
pages = {17},
isbn="978-1-4614-3673-7",
doi={10.1007/978-1-4614-3673-7_3},
prefix = {SoPh,31
275, 17},
url={https://doi.org/10.1007/978-1-4614-3673-7_3}
}

@article{kumar_x-ray_2015,
    title = {X-{Ray} and {EUV} {Observations} of {Simultaneous} {Short} and {Long} {Period} {Oscillations} in {Hot} {Coronal} {Arcade} {Loops}},
    volume = {804},
    issn = {0004-637X},
    url = {https://ui.adsabs.harvard.edu/abs/2015ApJ...804....4K},
    doi = {10.1088/0004-637X/804/1/4},
    urldate = {2023-08-08},
    journal = {The Astrophysical Journal},
    author = {Kumar, Pankaj and Nakariakov, Valery M. and Cho, Kyung-Suk},
    month = may,
    year = {2015},
    note = {ADS Bibcode: 2015ApJ...804....4K},
    pages = {4},
}

@ARTICLE{2019Nakariakov,
       author = {{Nakariakov}, V.~M. and {Kosak}, M.~K. and {Kolotkov}, D.~Y. and {Anfinogentov}, S.~A. and {Kumar}, P. and {Moon}, Y.-J.},
        title = "{Properties of Slow Magnetoacoustic Oscillations of Solar Coronal Loops by Multi-instrumental Observations}",
      journal = {\apjl},
         year = 2019,
        month = mar,
       volume = {874},
       number = {1},
          eid = {L1},
        pages = {L1},
          doi = {10.3847/2041-8213/ab0c9f},
       adsurl = {https://ui.adsabs.harvard.edu/abs/2019ApJ...874L...1N}
}

@ARTICLE{2002DeMoortel,
       author = {{De Moortel}, I. and {Ireland}, J. and {Hood}, A.~W. and {Walsh}, R.~W.},
        title = "{The detection of 3 \& 5 min period oscillations in coronal loops}",
      journal = {\aap},
         year = 2002,
        month = may,
       volume = {387},
        pages = {L13-L16},
          doi = {10.1051/0004-6361:20020436},
       adsurl = {https://ui.adsabs.harvard.edu/abs/2002A&A...387L..13D}
}

@ARTICLE{2006Marsh,
       author = {{Marsh}, M.~S. and {Walsh}, R.~W.},
        title = "{p-Mode Propagation through the Transition Region into the Solar Corona. I. Observations}",
      journal = {\apj},
         year = 2006,
        month = may,
       volume = {643},
       number = {1},
        pages = {540-548},
          doi = {10.1086/501450},
       adsurl = {https://ui.adsabs.harvard.edu/abs/2006ApJ...643..540M}
}

@ARTICLE{2015Prasad,
       author = {{Krishna Prasad}, S. and {Jess}, D.~B. and {Khomenko}, Elena},
        title = "{On the Source of Propagating Slow Magnetoacoustic Waves in Sunspots}",
      journal = {\apjl},
         year = 2015,
        month = oct,
       volume = {812},
       number = {1},
          eid = {L15},
        pages = {L15},
          doi = {10.1088/2041-8205/812/1/L15},
archivePrefix = {arXiv},
       eprint = {1510.03275},
 primaryClass = {astro-ph.SR},
       adsurl = {https://ui.adsabs.harvard.edu/abs/2015ApJ...812L..15K}
}

@ARTICLE{2025Tripathy,
       author = {{Tripathy}, Suraj K. and {Krishna Prasad}, S. and {Banerjee}, D.},
        title = "{Properties of Slow Magnetoacoustic Waves Observed Simultaneously Using Hi-C 2.1 and AIA}",
      journal = {\apj},
         year = 2025,
        month = jul,
       volume = {987},
       number = {1},
          eid = {57},
        pages = {57},
          doi = {10.3847/1538-4357/ade3c8},
archivePrefix = {arXiv},
       eprint = {2506.06126},
 primaryClass = {astro-ph.SR},
       adsurl = {https://ui.adsabs.harvard.edu/abs/2025ApJ...987...57T}
}

@ARTICLE{2025Baweja,
       author = {{Baweja}, Upasna and {Pant}, Vaibhav and {Krishna Prasad}, S. and {Shrivastav}, Arpit Kumar and {Van Doorsselaere}, Tom and {Narang}, Nancy and {Verbeeck}, Cis and {Khan}, M. Saleem and {Berghmans}, David},
        title = "{Coexistence of Longitudinal and Transverse Oscillations in Polar Plumes Observed with Solar Orbiter/Extreme Ultraviolet Imager}",
      journal = {\apjl},
         year = 2025,
        month = oct,
       volume = {991},
       number = {2},
          eid = {L45},
        pages = {L45},
          doi = {10.3847/2041-8213/ae071e},
archivePrefix = {arXiv},
       eprint = {2509.07796},
 primaryClass = {astro-ph.SR},
       adsurl = {https://ui.adsabs.harvard.edu/abs/2025ApJ...991L..45B}
}

@ARTICLE{2024Meadowcroft,
       author = {{Meadowcroft}, Rebecca L. and {Zhong}, Sihui and {Kolotkov}, Dmitrii Y. and {Nakariakov}, Valery M.},
        title = "{Observation of a propagating slow magnetoacoustic wave in a coronal plasma fan with SDO/AIA and SolO/EUI}",
      journal = {\mnras},
         year = 2024,
        month = jan,
       volume = {527},
       number = {3},
        pages = {5302-5310},
          doi = {10.1093/mnras/stad3506},
       adsurl = {https://ui.adsabs.harvard.edu/abs/2024MNRAS.527.5302M}
}

@ARTICLE{1998DeForest,
       author = {{DeForest}, C.~E. and {Gurman}, J.~B.},
        title = "{Observation of Quasi-periodic Compressive Waves in Solar Polar Plumes}",
      journal = {\apjl},
         year = 1998,
        month = jul,
       volume = {501},
       number = {2},
        pages = {L217-L220},
          doi = {10.1086/311460},
       adsurl = {https://ui.adsabs.harvard.edu/abs/1998ApJ...501L.217D}
}

@ARTICLE{1997Ofman,
       author = {{Ofman}, L. and {Romoli}, M. and {Poletto}, G. and {Noci}, G. and {Kohl}, J.~L.},
        title = "{Ultraviolet Coronagraph Spectrometer Observations of Density Fluctuations in the Solar Wind}",
      journal = {\apjl},
         year = 1997,
        month = dec,
       volume = {491},
       number = {2},
        pages = {L111-L114},
          doi = {10.1086/311067},
       adsurl = {https://ui.adsabs.harvard.edu/abs/1997ApJ...491L.111O}
}

@ARTICLE{2012KrishnaPrasad,
       author = {{Krishna Prasad}, S. and {Banerjee}, D. and {Van Doorsselaere}, T. and {Singh}, J.},
        title = "{Omnipresent long-period intensity oscillations in open coronal structures}",
      journal = {\aap},
         year = 2012,
        month = oct,
       volume = {546},
          eid = {A50},
        pages = {A50},
          doi = {10.1051/0004-6361/201219885},
archivePrefix = {arXiv},
       eprint = {1209.2536},
 primaryClass = {astro-ph.SR},
       adsurl = {https://ui.adsabs.harvard.edu/abs/2012A&A...546A..50K}
}

@ARTICLE{2010DePontieu,
       author = {{De Pontieu}, Bart and {McIntosh}, Scott W.},
        title = "{Quasi-periodic Propagating Signals in the Solar Corona: The Signature of Magnetoacoustic Waves or High-velocity Upflows?}",
      journal = {\apj},
         year = 2010,
        month = oct,
       volume = {722},
       number = {2},
        pages = {1013-1029},
          doi = {10.1088/0004-637X/722/2/1013},
archivePrefix = {arXiv},
       eprint = {1008.5300},
 primaryClass = {astro-ph.SR},
       adsurl = {https://ui.adsabs.harvard.edu/abs/2010ApJ...722.1013D}
}

@ARTICLE{2012Tian,
       author = {{Tian}, Hui and {McIntosh}, Scott W. and {Wang}, Tongjiang and {Ofman}, Leon and {De Pontieu}, Bart and {Innes}, Davina E. and {Peter}, Hardi},
        title = "{Persistent Doppler Shift Oscillations Observed with Hinode/EIS in the Solar Corona: Spectroscopic Signatures of Alfv{\'e}nic Waves and Recurring Upflows}",
      journal = {\apj},
         year = 2012,
        month = nov,
       volume = {759},
       number = {2},
          eid = {144},
        pages = {144},
          doi = {10.1088/0004-637X/759/2/144},
archivePrefix = {arXiv},
       eprint = {1209.5286},
 primaryClass = {astro-ph.SR},
       adsurl = {https://ui.adsabs.harvard.edu/abs/2012ApJ...759..144T}
}

@ARTICLE{2016Bryans,
       author = {{Bryans}, P. and {McIntosh}, S.~W. and {De Moortel}, I. and {De Pontieu}, B.},
        title = "{On the Connection between Propagating Solar Coronal Disturbances and Chromospheric Footpoints}",
      journal = {\apjl},
         year = 2016,
        month = sep,
       volume = {829},
       number = {1},
          eid = {L18},
        pages = {L18},
          doi = {10.3847/2041-8205/829/1/L18},
       adsurl = {https://ui.adsabs.harvard.edu/abs/2016ApJ...829L..18B}
}

@ARTICLE{2015Samanta,
       author = {{Samanta}, Tanmoy and {Pant}, Vaibhav and {Banerjee}, Dipankar},
        title = "{Propagating Disturbances in the Solar Corona and Spicular Connection}",
      journal = {\apjl},
         year = 2015,
        month = dec,
       volume = {815},
       number = {1},
          eid = {L16},
        pages = {L16},
          doi = {10.1088/2041-8205/815/1/L16},
archivePrefix = {arXiv},
       eprint = {1511.07354},
 primaryClass = {astro-ph.SR},
       adsurl = {https://ui.adsabs.harvard.edu/abs/2015ApJ...815L..16S}
}

@ARTICLE{2011Tian,
       author = {{Tian}, Hui and {McIntosh}, Scott W. and {De Pontieu}, Bart},
        title = "{The Spectroscopic Signature of Quasi-periodic Upflows in Active Region Timeseries}",
      journal = {\apjl},
         year = 2011,
        month = feb,
       volume = {727},
       number = {2},
          eid = {L37},
        pages = {L37},
          doi = {10.1088/2041-8205/727/2/L37},
archivePrefix = {arXiv},
       eprint = {1012.5112},
 primaryClass = {astro-ph.SR},
       adsurl = {https://ui.adsabs.harvard.edu/abs/2011ApJ...727L..37T}
}

@ARTICLE{2021BanerjeeSSRv,
       author = {{Banerjee}, D. and {Krishna Prasad}, S. and {Pant}, V. and {McLaughlin}, J.~A. and {Antolin}, P. and {Magyar}, N. and {Ofman}, L. and {Tian}, H. and {Van Doorsselaere}, T. and {De Moortel}, I. and {Wang}, T.~J.},
        title = "{Magnetohydrodynamic Waves in Open Coronal Structures}",
      journal = {\ssr},
         year = 2021,
        month = oct,
       volume = {217},
       number = {7},
          eid = {76},
        pages = {76},
          doi = {10.1007/s11214-021-00849-0},
archivePrefix = {arXiv},
       eprint = {2012.08802},
 primaryClass = {astro-ph.SR},
       adsurl = {https://ui.adsabs.harvard.edu/abs/2021SSRv..217...76B}
}

@ARTICLE{2002Aschwanden,
       author = {{Aschwanden}, Markus J. and {De Pontieu}, Bart and {Schrijver}, Carolus J. and {Title}, Alan M.},
        title = "{Transverse Oscillations in Coronal Loops Observed with TRACE   II. Measurements of Geometric and Physical Parameters}",
      journal = {\solphys},
         year = 2002,
        month = mar,
       volume = {206},
       number = {1},
        pages = {99-132},
          doi = {10.1023/A:1014916701283},
       adsurl = {https://ui.adsabs.harvard.edu/abs/2002SoPh..206...99A}
}

@ARTICLE{2009Milligan,
       author = {{Milligan}, Ryan O. and {Dennis}, Brian R.},
        title = "{Velocity Characteristics of Evaporated Plasma Using Hinode/EUV Imaging Spectrometer}",
      journal = {\apj},
         year = 2009,
        month = jul,
       volume = {699},
       number = {2},
        pages = {968-975},
          doi = {10.1088/0004-637X/699/2/968},
archivePrefix = {arXiv},
       eprint = {0905.1669},
 primaryClass = {astro-ph.SR},
       adsurl = {https://ui.adsabs.harvard.edu/abs/2009ApJ...699..968M}
}

@ARTICLE{2017Li,
       author = {{Li}, D. and {Ning}, Z.~J. and {Huang}, Y. and {Zhang}, Q.~M.},
        title = "{Explosive Chromospheric Evaporation Driven by Nonthermal Electrons around One Footpoint of a Solar Flare Loop}",
      journal = {\apjl},
         year = 2017,
        month = may,
       volume = {841},
       number = {1},
          eid = {L9},
        pages = {L9},
          doi = {10.3847/2041-8213/aa71b0},
archivePrefix = {arXiv},
       eprint = {1705.02448},
 primaryClass = {astro-ph.SR},
       adsurl = {https://ui.adsabs.harvard.edu/abs/2017ApJ...841L...9L}
}

@ARTICLE{2015Tian,
       author = {{Tian}, Hui and {Young}, Peter R. and {Reeves}, Katharine K. and {Chen}, Bin and {Liu}, Wei and {McKillop}, Sean},
        title = "{Temporal Evolution of Chromospheric Evaporation: Case Studies of the M1.1 Flare on 2014 September 6 and X1.6 Flare on 2014 September 10}",
      journal = {\apj},
         year = 2015,
        month = oct,
       volume = {811},
       number = {2},
          eid = {139},
        pages = {139},
          doi = {10.1088/0004-637X/811/2/139},
archivePrefix = {arXiv},
       eprint = {1505.02736},
 primaryClass = {astro-ph.SR},
       adsurl = {https://ui.adsabs.harvard.edu/abs/2015ApJ...811..139T}
}

@ARTICLE{2012Ofman,
       author = {{Ofman}, L. and {Wang}, T.~J. and {Davila}, J.~M.},
        title = "{Slow Magnetosonic Waves and Fast Flows in Active Region Loops}",
      journal = {\apj},
         year = 2012,
        month = aug,
       volume = {754},
       number = {2},
          eid = {111},
        pages = {111},
          doi = {10.1088/0004-637X/754/2/111},
archivePrefix = {arXiv},
       eprint = {1205.5732},
 primaryClass = {astro-ph.SR},
       adsurl = {https://ui.adsabs.harvard.edu/abs/2012ApJ...754..111O}
}

@INPROCEEDINGS{2012Wang,
       author = {{Wang}, T. and {Ofman}, L. and {Davila}, J.~M.},
        title = "{Propagating Intensity Disturbances in Fan-like Coronal Loops: Flows or Waves?}",
    booktitle = {4th Hinode Science Meeting: Unsolved Problems and Recent Insights},
         year = 2012,
       editor = {{Bellot Rubio}, L. and {Reale}, F. and {Carlsson}, M.},
       series = {Astronomical Society of the Pacific Conference Series},
       volume = {455},
        month = may,
        pages = {227},
          doi = {10.48550/arXiv.1101.6017},
archivePrefix = {arXiv},
       eprint = {1101.6017},
 primaryClass = {astro-ph.SR},
       adsurl = {https://ui.adsabs.harvard.edu/abs/2012ASPC..455..227W}
}

@ARTICLE{2011Nishizuka,
       author = {{Nishizuka}, N. and {Hara}, H.},
        title = "{Spectroscopic Observations of Continuous Outflows and Propagating Waves from NOAA 10942 with Extreme Ultraviolet Imaging Spectrometer/Hinode}",
      journal = {\apjl},
         year = 2011,
        month = aug,
       volume = {737},
       number = {2},
          eid = {L43},
        pages = {L43},
          doi = {10.1088/2041-8205/737/2/L43},
       adsurl = {https://ui.adsabs.harvard.edu/abs/2011ApJ...737L..43N}
}

@ARTICLE{2019Kolotkov,
       author = {{Kolotkov}, D.~Y. and {Nakariakov}, V.~M. and {Zavershinskii}, D.~I.},
        title = "{Damping of slow magnetoacoustic oscillations by the misbalance between heating and cooling processes in the solar corona}",
      journal = {\aap},
         year = 2019,
        month = aug,
       volume = {628},
          eid = {A133},
        pages = {A133},
          doi = {10.1051/0004-6361/201936072},
archivePrefix = {arXiv},
       eprint = {1907.07051},
 primaryClass = {astro-ph.SR},
       adsurl = {https://ui.adsabs.harvard.edu/abs/2019A&A...628A.133K}
}

@ARTICLE{2026Paliwal,
       author = {{Paliwal}, Hitesh and {Krishna Prasad}, S. and {Krishna}, M.~V. Sunil},
        title = "{Sloshing Oscillations in coronal loops excited by successive M- and C-Class flares}",
      journal = {arXiv e-prints},
         year = 2026,
        month = jul,
          eid = {arXiv:2607.17685},
        pages = {arXiv:2607.17685},
          doi = {10.48550/arXiv.2607.17685},
archivePrefix = {arXiv},
       eprint = {2607.17685},
 primaryClass = {astro-ph.SR},
       adsurl = {https://ui.adsabs.harvard.edu/abs/2026arXiv260717685P}
}

@ARTICLE{2024Karakotov,
       author = {{Karakotov}, Ruslan and {Kuznetsov}, Alexey and {Anfinogentov}, Sergey and {Nakariakov}, Valery M.},
        title = "{Sloshing Oscillations in Hot Coronal Loops Associated with M-Class Flares}",
      journal = {\solphys},
         year = 2024,
        month = dec,
       volume = {299},
       number = {12},
          eid = {163},
        pages = {163},
          doi = {10.1007/s11207-024-02404-w},
       adsurl = {https://ui.adsabs.harvard.edu/abs/2024SoPh..299..163K}
}

@ARTICLE{2023Cozzo,
       author = {{Cozzo}, Gabriele and {Pagano}, Paolo and {Petralia}, Antonino and {Reale}, Fabio},
        title = "{Asymmetric Twisting of Coronal Loops}",
      journal = {Symmetry},
         year = 2023,
        month = mar,
       volume = {15},
       number = {3},
          eid = {627},
        pages = {627},
          doi = {10.3390/sym15030627},
       adsurl = {https://ui.adsabs.harvard.edu/abs/2023Symm...15..627C}
}

@ARTICLE{2013Wang,
       author = {{Wang}, Tongjiang and {Ofman}, Leon and {Davila}, Joseph M.},
        title = "{Three-dimensional Magnetohydrodynamic Modeling of Propagating Disturbances in Fan-like Coronal Loops}",
      journal = {\apjl},
         year = 2013,
        month = sep,
       volume = {775},
       number = {1},
          eid = {L23},
        pages = {L23},
          doi = {10.1088/2041-8205/775/1/L23},
archivePrefix = {arXiv},
       eprint = {1308.0282},
 primaryClass = {astro-ph.SR},
       adsurl = {https://ui.adsabs.harvard.edu/abs/2013ApJ...775L..23W}
}

@ARTICLE{2019Simoes,
       author = {{Sim{\~o}es}, Paulo J.~A. and {Reid}, Hamish A.~S. and {Milligan}, Ryan O. and {Fletcher}, Lyndsay},
        title = "{The Spectral Content of SDO/AIA 1600 and 1700 {\r{A}} Filters from Flare and Plage Observations}",
      journal = {\apj},
         year = 2019,
        month = jan,
       volume = {870},
       number = {2},
          eid = {114},
        pages = {114},
          doi = {10.3847/1538-4357/aaf28d},
archivePrefix = {arXiv},
       eprint = {1808.01488},
 primaryClass = {astro-ph.SR},
       adsurl = {https://ui.adsabs.harvard.edu/abs/2019ApJ...870..114S}
}
\bibliographystyle{aasjournal}

\clearpage
\appendix 
\section{Time-distance maps and duration of plasma injection for Events II, III and IV} \label{appendix:A}

\begin{figure*}[!htbp]
    \centering
    \includegraphics[width=\linewidth,trim=0.8cm 0cm 0.8cm 0cm]{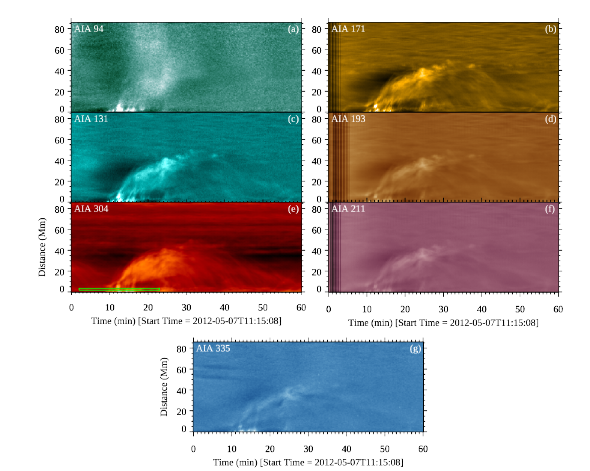}
    \caption{Similar to Figure \ref{fig:fig2_1900}, but corresponding to the artificial slit depicted in Figure \ref{fig:fig1_context}(b).}
    \label{fig:appendix_fig1_1045}
\end{figure*}

\begin{figure*}
    \centering
    \includegraphics[width=\linewidth,trim=0.8cm 0cm 0.8cm 0.1cm]{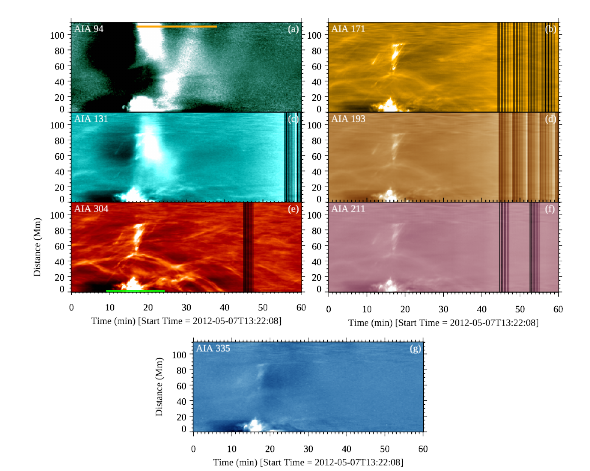}
    \caption{Similar to Figure \ref{fig:fig2_1900}, but corresponding to the artificial slit depicted in Figure \ref{fig:fig1_context}(c).}
    \label{fig:appendix_fig2_1300}
\end{figure*}

\begin{figure*}
    \centering
    \includegraphics[width=\linewidth,trim=0.8cm 0cm 0.8cm 0.1cm]{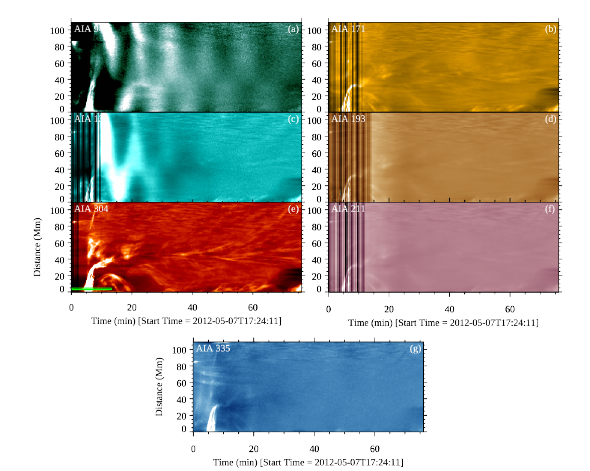}
    \caption{Similar to Figure \ref{fig:fig2_1900}, but corresponding to the artificial slit depicted in Figure \ref{fig:fig1_context}(d).}
    \label{fig:appendix_fig3_1700}
\end{figure*}

\begin{figure*}
    \centering
    \includegraphics[width=\linewidth,trim=0cm 0cm 0cm 0cm]{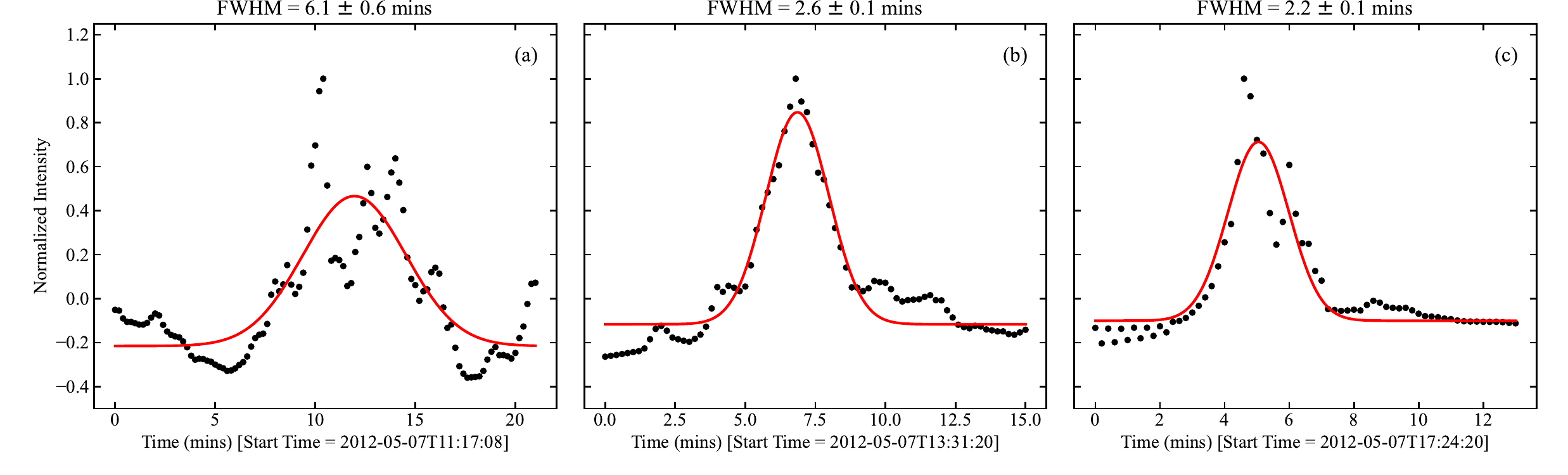}
    \caption{The scatter points in panels (a), (b), and (c) represent the temporal variation of normalized intensity corresponding to the green boxes shown in Figure \ref{fig:appendix_fig1_1045} (e), Figure \ref{fig:appendix_fig2_1300} (e), and \ref{fig:appendix_fig3_1700} (e). The red curve shows the fitted Gaussian function on these scatter points.}
    \label{fig:appendix_fig4_flare_duration}
\end{figure*}

\end{document}